\documentclass[12pt]{article}

\usepackage{graphics,cite,amssymb,epsfig,float,psfrag}
\usepackage[usenames,dvips]{color}
\usepackage{rotating}
\newcommand{\lsim}{\raisebox{-0.13cm}{~\shortstack{$<$ \\[-0.07cm] $\sim$}}~} 
\newcommand{\gsim}{\raisebox{-0.13cm}{~\shortstack{$>$ \\[-0.07cm] $\sim$}}~} 
\newcommand{\beq}{\begin{eqnarray}} 
\newcommand{\eeq}{\end{eqnarray}} 

\newcommand{\s}{\\ \vspace*{-4mm}}

\begin{document}

\vspace{.8cm}

\hfill  LPT--ORSAY--12--65

\hfill CERN--PH--TH/2012--187

\hfill KA-TP-27-2012

\hfill SFB/CPP-12-46

\vspace*{1.4cm}

\begin{center}

{\large\bf The apparent excess in the Higgs to di-photon rate}
 
\vspace*{2mm} 

{\large\bf  at the LHC: New Physics or QCD uncertainties?}

\vspace*{.6cm}

{\sc J. Baglio$^1$, A. Djouadi$^{2,3}$ and R.M. Godbole$^{3,4}$} 

\vspace*{.8cm}

\begin{small}

$^1$ Institut f\"ur Theoretische Physik, KIT,  D-76128 Karlsruhe, Germany.  

\vspace*{1mm}

$^2$ Laboratoire de Physique Th\'eorique, U. Paris--Sud and  CNRS,  F--91405
Orsay, France.

\vspace*{1mm}

$^3$ Theory Unit, Department of Physics, CERN, CH-1211 Geneva 23, 
Switzerland.

\vspace*{1mm}

$^4$ Center for High Energy Physics, Indian Institute of Science, Bangalore 560 012, 
India.

\end{small}

\end{center}

\vspace*{1cm}

\begin{abstract} 

The Higgs boson with a mass $M_H \approx 126$ GeV has been observed by
the ATLAS and CMS experiments at the LHC and a total significance of
about five standard deviations has been reported by both
collaborations when the channels $H\to \gamma \gamma$ and $H\to ZZ \to
4\ell$ are combined.  Nevertheless, while the rates in the later
search channel appear to be in accord with those predicted in the
Standard Model, there seems to be an excess of data in the case of
the $H\to \gamma\gamma$ discovery channel. Before invoking new physics
contributions to explain this excess in the di--photon Higgs rate, one
should verify that standard  QCD effects cannot account for it. We
describe how the theoretical uncertainties in the Higgs boson cross
section for the main production process at the LHC, $gg \to H$, which
are known to be large, should be incorporated in practice. We further
show that the  discrepancy between the theoretical prediction and the
measured value of the $gg \!\to \!H \! \to \! \gamma \gamma$ rate,
reduces to about one standard deviation when the QCD uncertainties are
taken into account.

\end{abstract} 

\newpage

The Higgs particle \cite{Higgs} has been, at last, observed by the
ATLAS and CMS experiments at the LHC as a signal with about five
standard deviations has been reported by each collaboration when the
main search channels are combined \cite{LHC}. This discovery
represents a triumph for the Standard Model (SM) of particle physics
and crowns more than four decades of theoretical and experimental
endeavour. Now that the Higgs discovery chapter is closing, a new and
even more challenging chapter is opening: the verification of the
fundamental properties of the particle and the precise determination
of its couplings. This program can be started  at the LHC since, for
the reported mass $M_H \approx 126$ GeV \cite{LHC}, one can have
access to the Higgs boson in many production and decay channels
\cite{Review}.

At the LHC, the main Higgs production channel is the top and bottom
quark loop mediated gluon--gluon fusion mechanism $gg \to H$: at
center of mass energies of $\sqrt s=7$ and 8 TeV and for a Higgs boson
mass of $M_H\approx 126$ GeV, the inclusive cross sections are about
$\sigma(gg\to H)\approx 15$ pb and 20 pb, respectively
\cite{LHCXS,BD}. The vector boson fusion $qq \to Hqq$ and the
Higgs--strahlung $q\bar q \to HW+HZ$ mechanisms add only little to
these rates, respectively $\approx 8\%$ and $\approx 5.5\%$, before
kinematical cuts are applied \cite{LHCXS,BD}. The $gg \to H$ cross
section is known up to next--to--next--to--leading order (NNLO) in
perturbative QCD: the $K$--factor defined as the ratio of the higher
order to the leading order (LO) \cite{ggH-LO} cross sections is
$\approx 1.8 $ at NLO \cite{ggH-NLO1,ggH-NLO2} and $\approx 2.5$ at
NNLO \cite{ggH-NNLO}. The cross section receives also small
contributions from the resummation of soft gluons \cite{ggH-NNLL} and
electroweak corrections \cite{ggH-EW, ggH-mix}. Some small corrections
that go beyond NNLO accuracy are also available \cite{ggH-NNNLO} but
have not been included in the predictions used by the LHC experimental
collaborations. It is clear that it is this exceptionally large
$K$--factor that allows for a sensitivity to the Higgs boson at the
LHC with the presently collected data. The main Higgs search channels
take advantage of the clean $H\! \to \! \gamma \gamma$,  $H \! \to \!
ZZ \! \to \! 4\ell^\pm$ and $H \! \to \! WW \! \to \! \ell \nu \ell
\nu$ final states (with $\ell\!=\!e,\mu$), while the significance of
other modes such as $H\! \to\! \tau^+\tau^-$ and $VH \! \to \! V b\bar
b$ is presently low. The Higgs decay branching ratios are rather
precisely known \cite{HDECAY}.

The results presented by the ATLAS and CMS Collaborations on their
Higgs search \cite{LHC} turn out to be a little surprising. Indeed,
while the rates for the $H\to ZZ$ search channel seem to reasonably
agree with the SM expectations, a discrepancy mostly driven by the
ATLAS results is observed in the channel $H\to \gamma \gamma$ which
has the largest signal significance. Defining the ratios $R_{\rm
  XX}=\sigma^{\rm obs}_{\rm H\to XX}/ \sigma^{\rm SM}_{\rm H\to XX}$
of the measured cross section in a given search channel compared to
the theoretical expectation, one finds for the two
experiments\footnote{The numbers given by the collaborations
  correspond to the optimal (i.e. which maximises the likelihood of
  the test statistics) value $\hat \mu$ of the signal strength
  modifier that multiplies the expected cross section such that $\hat
  \mu = \sigma/\sigma^{\rm SM}$. As such, strictly speaking, they are
  not the true cross section ratios $R_{XX}$. We will nevertheless
  assume that they are the same for simplicity and for purposes of
  illustrating our point.}
\begin{equation} 
\begin{tabular}{lll}
ATLAS: & $R_{\gamma \gamma}=1.90 \pm 0.50\, ,$  & $R_{ZZ}=1.3 \pm 0.6 \, ,$  \\
CMS: & $R_{\gamma \gamma}=1.56 \pm 0.43 \, , $  & $R_{ZZ}=0.7 \pm 0.5 \, ,$ \\
ATLAS$\oplus$CMS: & $R_{\gamma \gamma}=1.71 \pm 0.33 \, , $  & $R_{ZZ}=0.95 
\pm 0.40 \, .$ 
\end{tabular}
\label{masterequation}
\end{equation}
It then seems that the rate in the $H\to \gamma \gamma$ channel is
more than two standard deviations larger compared to the SM
prediction, when the ATLAS and CMS measurements are combined. This is
a rather exciting situation as, not only the Higgs boson has been
finally discovered, but in addition it appears to come with hints of
some new physics. This could be the first (long awaited) signal for
beyond the SM physics at the LHC. It is expected that a large number
of studies (including some by the present authors and in addition to
the many which were done after the first hint for this excess was
reported in the 2011 data) will be devoted to the explanation of this
feature in terms of new phenomena\footnote{However, because $R_{ZZ}$
  seems to be in agreement with the SM and there is no sign of a new
  particle in direct searches at the LHC, the $H\to \gamma\gamma$
  excess will be particularly difficult to accommodate in relatively
  simple and/or well motivated SM extensions. One would probably have
  to resort to slightly ``baroque" new constructions or scenarios to
  explain the excess.}.

However, before doing so, it would be probably wiser to consider more
conventional explanations for this intriguing excess. The first one
would be simply that this is the result of a statistical fluctuation
(in both signal and backgrounds); after all, many $\gsim 2\sigma$
excesses appeared in the recent years and faded away when more data
was collected\footnote{It might be noted that history could repeat
  itself: in the first $Z\to \ell^+\ell^-$ events observed by both the
  UA1 and UA2 Collaborations and which led to the discovery of the $Z$
  boson in 1983, a significant fraction were accompanied by additional
  photons \cite{UA1}. This triggered a plethora of papers proposing
  composite models of quarks, leptons and weak bosons, before the
  excess of photons died away with more statistics.}.
Another possibility would be that the systematical uncertainties
in the extremely difficult $H\to \gamma \gamma$ channel have been
underestimated\footnote{In particular, a significant fraction of the
  $\gamma \gamma$ events seems to come with two additional jets, while
  the predicted rate in the SM from vector boson fusion $qq\to qqH$
  and $gg \to Hgg$ is expected to be small.}; potential experimental
problems (if any) could also be fixed when more data is accumulated
and the detector response better understood.

A third conventional possibility to explain the $H\to \gamma \gamma$
excess would be that the QCD uncertainties may have been
underestimated\footnote{An example out of many for such a possibility
  is the $p\bar p \to b\bar b$ production cross section at the
  Tevatron that had been first determined to be a factor of two to
  three larger than the QCD prediction, before higher order effects
  and various uncertainties were included; see Ref.~\cite{Foot0} for a
  discussion.} by the experimental collaborations. This is the option
that we will investigate in the present paper. We will show that if
the theoretical uncertainties in the prediction of the cross section
for the by far dominant $gg\to H$ production process at the LHC are
properly included, the significance of the di-photon excess becomes
substantially lower and agreement between theory and experiment can be
reached at the $\approx 1 \sigma$ level. This is achieved at the
expense of slightly increasing the discrepancy of the $H\to ZZ$ and
$H\to WW$ signals which, however, are affected by much larger
experimental (mainly statistical) uncertainties.\smallskip

Let us start by discussing the two main theoretical uncertainties that
enter into play in $\sigma(gg\to H)$ and which have been discussed in
detail by the LHC Higgs cross section working group (LHCHWG)
\cite{LHCXS} and in Refs.~\cite{BD,BD1,BDG,Anastasiou,Grazzini}: the
scale and PDF+$\alpha_s$ uncertainties\footnote{An addendum with the
  complete analysis for Higgs production at $\sqrt s=8$ TeV has been
  added to the version of Ref.~\cite{BD} submitted to the archives
  (arXiv:1012.0530v5).}.

The perturbative QCD corrections to the $gg\to H$ cross section are so
large, leading to a $K$--factor of about 2.5, that it raises worries
about the rate of convergence of the perturbative series. The
possibility of still large higher order contributions beyond NNLO
hence cannot be totally excluded. The effects of the unknown
contributions are usually estimated from the variation of the cross
section with the renormalisation $\mu_R$ and factorisation $\mu_F$
scales at which the process is evaluated. In the $gg\to H$ process,
the median scale is taken to be $\mu_R\!=\!\mu_F\!=\!\mu_0\! =\!
\frac12 M_H$ \cite{BD,Anastasiou} in order to absorb some of the
soft--gluon resummation corrections. Indeed, $\sigma(gg\to H)$
calculated at NNLO with $\mu_0=\frac12 M_H$ is then approximately the
same (up to a few percent) as the resummed cross section at
next-next-to-leading-logarithm (NNLL) with $\mu_0=M_H$
\cite{Grazzini}. However, the scale variation of the two cross
sections is different.

To estimate the scale uncertainty, the current convention is to vary
the renormalisation and factorisation scales within the range
$\mu_0/\kappa \!\le \!\mu_R, \mu_F \!\le \! \kappa \mu_0$, with the
ratio of scales restricted to the range $1/\kappa  \leq \mu_F/\mu_R
\leq \kappa$. The choice $\kappa = 2$ is usually adopted. For a Higgs
mass $M_H=126$ GeV, this leads to a scale uncertainty of $\Delta
\sigma_\mu \approx ^{+9\%}_{-10\%}$ at $\sqrt{s}= 7$ TeV and $\Delta
\sigma_\mu \approx ^{+12\%}_{-9.5\%}$ at $\sqrt {s}=8$ TeV when the
constraint $\frac12 \leq \mu_F/\mu_R \leq 2$ is imposed\footnote{The
  uncertainty is a few \% smaller if the two scales are equated,
  $\mu_R=\mu_F$, and one would obtain at NNLO $\Delta \sigma_\mu
  \approx ^{+8.7\%}_{-9.5\%}$  at $\sqrt{s}=8$ TeV in accord with
  Ref.~\cite{BD,Anastasiou}. The small difference for the central
  value of the cross section, obtained in our case by using the latest
  version of the program {\tt HIGLU} \cite{Higlu}, is due to some
  refinements in the treatment of the electroweak corrections
  performed in Ref.~\cite{Anastasiou}.}. Slightly larger uncertainties
occur if the scale is varied in a wider domain. If for instance, if
one chooses $\kappa=3$, the scale variation  would lead to a few
percent more uncertainty.

A second issue is related to the not yet entirely satisfactory
determination of the parton distribution functions (PDFs) and in
particular, the gluon densities. In addition to this, since
$\sigma^{\rm LO}(gg\!\to \!H) \propto \alpha_s^2$ and receives large
contributions at ${\cal O}(\geq \alpha_s^3)$, a modest change of
$\alpha_s$ (which is possible as the average $\alpha_s$ value
\cite{PDG} can be rather different from the ones obtained from
deep-inelastic scattering data that are used in some of the PDFs) can
lead to sizable change in the cross section value. There is a
statistical method to estimate the PDF uncertainties by allowing a
1$\sigma$ (or more) excursion of the experimental data that is used to
perform the global fits. In addition, the MSTW Collaboration
\cite{PDF-MSTW} provides a scheme that allows for a combined
evaluation of the PDF and $\alpha_s$ uncertainties. 

To take into account this additional uncertainty and the spread in the
predictions using the various NNLO PDF sets \cite{PDF-MSTW,PDFs}, the
PDF4LHC working group recommends~\cite{PDF4LHC} to take as a global
PDF uncertainty the MSTW PDF+$\Delta^{\rm exp}\alpha_s$ uncertainty at
the 68\% confidence level (CL) and multiply it by a factor of
two. This procedure gives nearly the same answer as the one proposed
in Refs.~\cite{BD1,BDG} in which one evaluates the combined 90\% CL
MSTW PDF+$\Delta^{\rm exp}\alpha_s+\Delta^{\rm th}\alpha_s$
uncertainty, where $\Delta^{\rm th}\alpha_s$ is for the error
generated by the theoretical uncertainty on $\alpha_s$ estimated to be
$\approx 0.002$ at NNLO \cite{PDF-MSTW}. For $\sigma(gg\to H)$ at
NNLO, one then finds a total PDF uncertainty of $\Delta\sigma^{\rm
  PDF} \approx 9\%$ at both $\sqrt s=7$ and 8 TeV for a $126$ GeV
Higgs boson.

\begin{figure}[hbtp]
\begin{center}
\mbox{
\epsfig{file=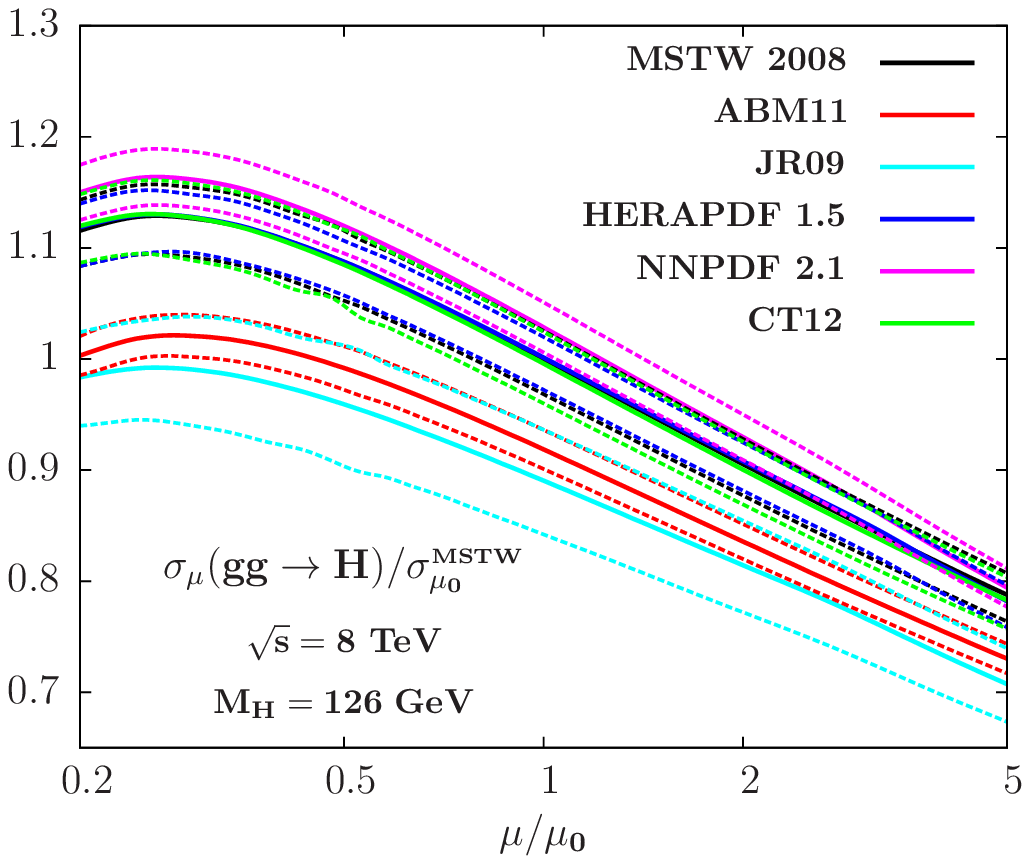,scale=0.73}\hspace*{5mm} 
\epsfig{file=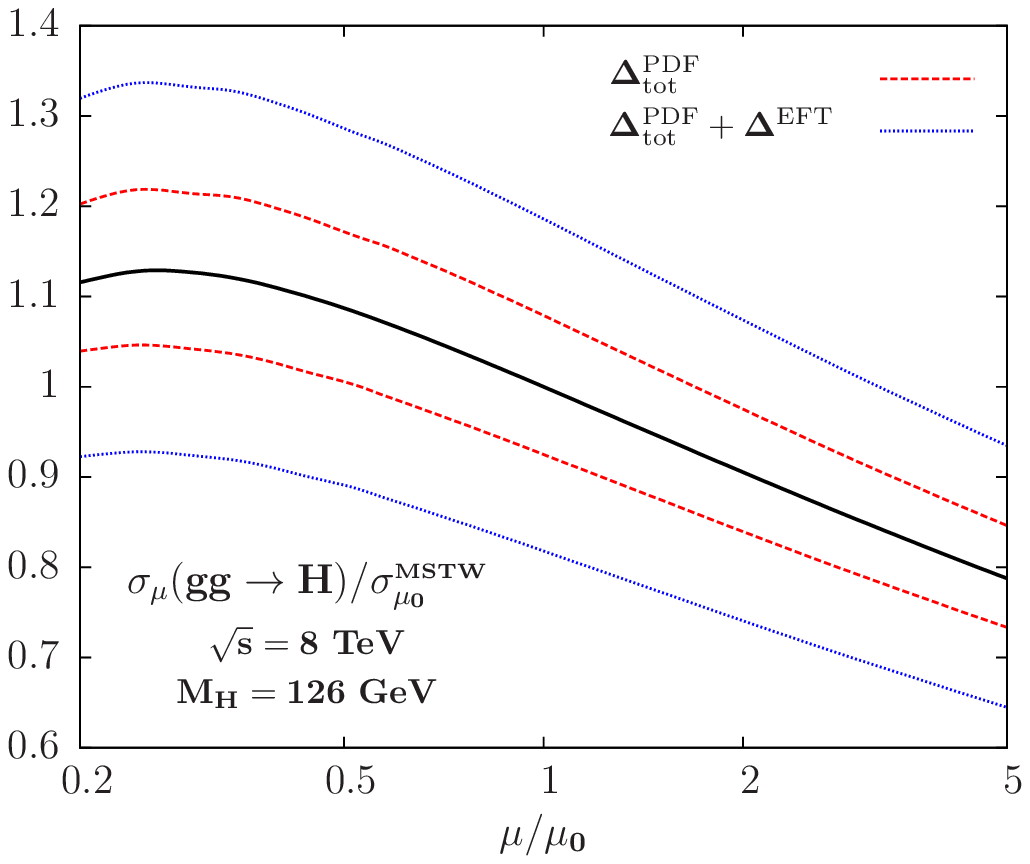,scale=0.73} }
\end{center}
\vspace*{-5mm}
\caption[]{\small $\sigma(gg\to H)$ at NNLO as a function of
  $\mu/\mu_0$ for $M_H=126$ GeV and $\sqrt s=8$ TeV. Left: the spread
  in the cross sections when the six NNLO PDF sets are used,
  normalised to the central MSTW cross section with $\mu_0= \frac12
  M_H$. Right: the relative PDF+$\alpha_s$ uncertainties in the MSTW
  case, when compared to the central value, as advocated by the LHCHWG
  group as well as the total uncertainty when the EFT uncertainty is
  linearly added.}
\vspace*{-1mm}
\label{PDFs}
\end{figure}

This discussion is illustrated in Fig.~\ref{PDFs} where the $gg\to H$
inclusive cross section for a $M_H=126$ GeV Higgs boson, evaluated at
NNLO--QCD and including the electroweak corrections, is displayed as a
function of the reduced scale $\mu/\mu_0$ with $\mu=\mu_R= \mu_F$ and
$\mu_0=\frac12 M_H$ at $\sqrt s=8$ TeV. The situation at $\sqrt s=7$
TeV is very similar. One can see that indeed, for the usual choice
$\mu/\mu_0=\frac12$ ($2$) of scale variation, the cross section
increases (decreases) by $\approx 10\%$. If one is conservative and
enlarges the domain of scale variation,  one notices that
$\sigma(gg\to H)$ decreases monotonically with increasing  $\mu/\mu_0$
but it has  a plateau for about $\mu/\mu_0\approx \frac13$ where the
cross section is maximal.

In the left-hand side of Fig.~\ref{PDFs}, the spread in the prediction
for $\sigma(gg\to H)$ is shown for the six PDF sets that are available
at NNLO \cite{PDF-MSTW,PDFs}, including the pure 90\%CL PDF
uncertainty bands. One sees that the spread of the cross sections is
rather significant, but most sets predict a rate that is smaller than
the MSTW prediction except for NNPDF and HERAPDF. The right--hand side
of the figure shows the PDFs uncertainty that the PDF4LHC group
\cite{PDF4LHC} recommends to retain, i.e. the MSTW  PDF$+\Delta^{\rm
  exp}\alpha_s$ at 68\%CL combined uncertainty multiplied by a factor
of two, which is about $\pm 9\%$ in the entire $\mu/\mu_0$ range.

A critical issue is the way the scale and PDF uncertainties should be
combined. As advocated by the LHCHWG \cite{LHCXS}, one should be
conservative and add the two uncertainties linearly; this is
equivalent of assuming that the PDF uncertainty is a pure theoretical
uncertainty with a flat prior\footnote{Despite the fact that the
  Hessian method provides an error that is of probabilistic nature, it
  does not account for the theoretical assumptions that enter into the
  PDF parametrisation. This theoretical uncertainty is reflected in
  the larger spread in the central values of the PDF
  predictions.}. For  $M_H\approx 126$ GeV, this procedure leads  to a
total uncertainty of about $\Delta^{\rm scale+PDF}\sigma\approx \pm
20\%$. This is shown in the right--hand side of Fig.~\ref{PDFs}.\s

There is, however, a third source source of uncertainty which has not
been accounted for by the LHCHWG~\cite{LHCXS} but has been discussed
in Ref.~\cite{BD,BD1,BDG}.  As the gluon--gluon fusion  process,
already at LO, occurs at the one--loop level with the additional
complication of having to account for the finite mass of the loop
particle, the NLO calculation is extremely complicated and the NNLO
calculation a formidable task. Luckily, one can work in an effective
field theory (EFT) approach in which the heavy quark in the loop is
integrated out, making the calculation of the contributions beyond NLO
possible. While this approach is justified for the dominant top quark
contribution for $M_H\! \lsim\! 2m_t$ \cite{EFT-mt}, it is not valid
for the $b$-quark loop  (and for the interference between the $b$--
and the $t$--loops) and for those involving the electroweak gauge
bosons \cite{ggH-EW}. The uncertainties induced by the use of the EFT
approach at NNLO are estimated, from the NLO case in which both the
exact and EFT calculations are available, to be of ${\cal O}(9\%)$ for
$M_H\approx 126$ GeV \cite{BD}.

This uncertainty is of pure theoretical origin as it is due to an
approximation in the calculation and has nothing to do with the scale
uncertainty. Hence it should be added linearly to the scale+PDF
uncertainty. For $M_H\approx 126$ GeV, this leads to a total
uncertainty of $\Delta^{\rm  tot} \sigma \approx \pm 30\%$ on the NNLO
$gg\to H$ cross section when the scale is varied in the commonly
adopted range $\frac12 \leq \mu/\mu_0 \leq 2$ as also shown in the
right--hand side of Fig.~\ref{PDFs}. 

In the case of interest,  i.e. for the normalised cross section times
branching ratios $R_{XX}$ given by the ATLAS and CMS Collaborations,
the story is not yet over and there are in fact two additional sources
which lead to uncertainties or normalisation problems, albeit smaller
than the ones discussed above:

$i)$ There are first uncertainties in the Higgs branching
ratios. Indeed, while Higgs decays into leptons and gauge bosons are
well under control, as mainly small electroweak effects are involved,
the partial widths into quark pairs and gluons are subject to
uncertainties. These are mainly due to the errors on input values of
the bottom and charm quark masses and $\alpha_s$, which then migrate
to the other decays branching fractions \cite{BD,BRs-MS}. For
$M_H\approx 126$ GeV, one would obtain $\Delta {\rm
  BR}(\gamma\gamma,WW,ZZ)\approx \pm 2\%$ \cite{BD} when slightly
smaller errors on the input masses $m_b$ and $m_c$ \cite{BRs-MS}
compared to the PDG input values \cite{PDG} are adopted.

$ii)$ There is a slight problem with the overall normalisation of
$\sigma(gg\to H)$. The normalisation adopted by the experiments (and
which comes from the LHCHWG) is the one obtained at NNLL
\cite{Grazzini} and not at NNLO \cite{BD,Anastasiou}. Besides the fact
that it is theoretically not entirely consistent to use the resummed
result (as the PDFs are defined at NNLO and not NNLL), the resummation
is available only for the inclusive rate and not for the cross
sections when experimental cuts are incorporated and that are actually
used by the experiments. It turns out that for $M_H\approx 126$ GeV,
$\sigma^{\rm NNLL}$ is $\approx 3\%$ smaller than $\sigma^{\rm NNL0}$
and has a smaller scale dependence \cite{Grazzini}. Hence, the $gg \to
H$ cross section might have been underestimated from the very
beginning, albeit by only a small amount.

These might affect the total rate and the uncertainties, increasing
them by a few percent. If one also takes a non-dogmatic approach and
increase the scale variation beyond the commonly chosen range $\frac12
\leq \mu/m_0 \leq 2$, one might end up with a total theoretical
uncertainty that is closer to $\Delta^{\rm th}  \sigma  \approx 40\%$
than 30\% for $M_H \approx 126$ GeV.

We are now in a position to discuss the impact of this total
uncertainty on the rates for Higgs production times decay branching
ratios in the channels $H\to \gamma \gamma$ and $H\to ZZ$ that have
been measured by the ATLAS and CMS Collaborations. A very important
fact to note from the very beginning is that, in the experimental
combination of different uncertainties, the theoretical uncertainty is
not treated as a bias, as should be the case\footnote{Let us
  illustrate this important point by calculating $\sigma(gg\to H)$ in
  a consistent way but different from the one which gives the central
  value of Ref.~\cite{LHCXS} and which has been adopted as a
  normalisation by the ATLAS and CMS Collaborations. We choose to use
  the NNPDF2.1 set and evaluate the cross section at a scale
  $\mu_R=\mu_F=\mu_0=\frac14 M_H$, which is within the range adopted
  for the scale uncertainty. We then obtain a NNLO cross section of
  $\sigma(gg\to H)= 22.9$ pb for $M_H=126$ GeV and $\sqrt{s}=8$
  TeV. This value is $\approx 20\%$ larger than the reference value of
  $\sigma(gg\to H)=19.2$ pb used by ATLAS and CMS. It is therefore
  clear that treating this theoretical uncertainty, which leads to the
  20\% change in the normalisation, as a mere nuisance can affect the
  conclusions in a very significant way.} but, instead, as a nuisance
parameter. Therefore, the scale and PDF uncertainties are not added
linearly to the experimental uncertainties in contradiction with the
LHCHWG recommendation, but are combined quadratically with the
experimental statistical and systematical errors. As the latter are
much larger than the scale and PDF uncertainties, at least 30\% for
the experimental errors and only 10\% for the scale and 10\% for the
PDF uncertainties, the magic of statistics and the combination in
quadrature makes that 
\beq \Delta^{\rm tot}  \sigma = \sqrt{
  (\Delta^{\rm exp}  \sigma)^2 +   (\Delta^\mu   \sigma)^2 +
  (\Delta^{\rm PDF}  \sigma)^2 } \approx 
\Delta^{\rm exp}   \sigma ~~{\rm for}~~\Delta^{\rm exp}  \sigma \gg
\Delta^\mu  \sigma,  \Delta^{\rm PDF}  \sigma \nonumber  \eeq 
This means that for $\Delta^{\rm exp}  \sigma \approx 30\%$, which is
the minimal experimental error, one would obtain only $\Delta^{\rm
  tot} \sigma  \approx 33\%$. Hence, one can consider that, in
practice, the above mentioned theoretical uncertainties are simply not
reflected in the errors of eq.~(\ref{masterequation}) for the ATLAS
and CMS measurements of the rates in the different channels.

For the comparison of theoretical predictions and the experimental
measurements, it is more convenient to adopt the procedure advocated,
for instance, in Ref.~\cite{BDG}, that is to ignore the theoretical
uncertainties in the likelihood fit performed in the experimental
analyses and simply confront the pure experimental error with the
theoretical prediction that includes the theory uncertainty band. For
the cross section in a given channel\footnote{Although the situation
  here might be slightly complicated as, in practice, one has to add
  to the cross section of the $gg \to H$ process those of the vector
  boson fusion and Higgs--strahlung processes. However, because the
  $gg\to H$ cross section is an order of magnitude larger than the
  summed cross section of the vector boson and Higgs-strahlung
  processes, the latter can be omitted  in a first approximation.} and
because the experimental values obtained from the multivariate
analyses have been cross-checked by a cut based analysis, it should be
possible to use such a procedure.

\begin{figure}[hbtp]
\vspace*{-.2cm}
\begin{center}
\epsfig{file=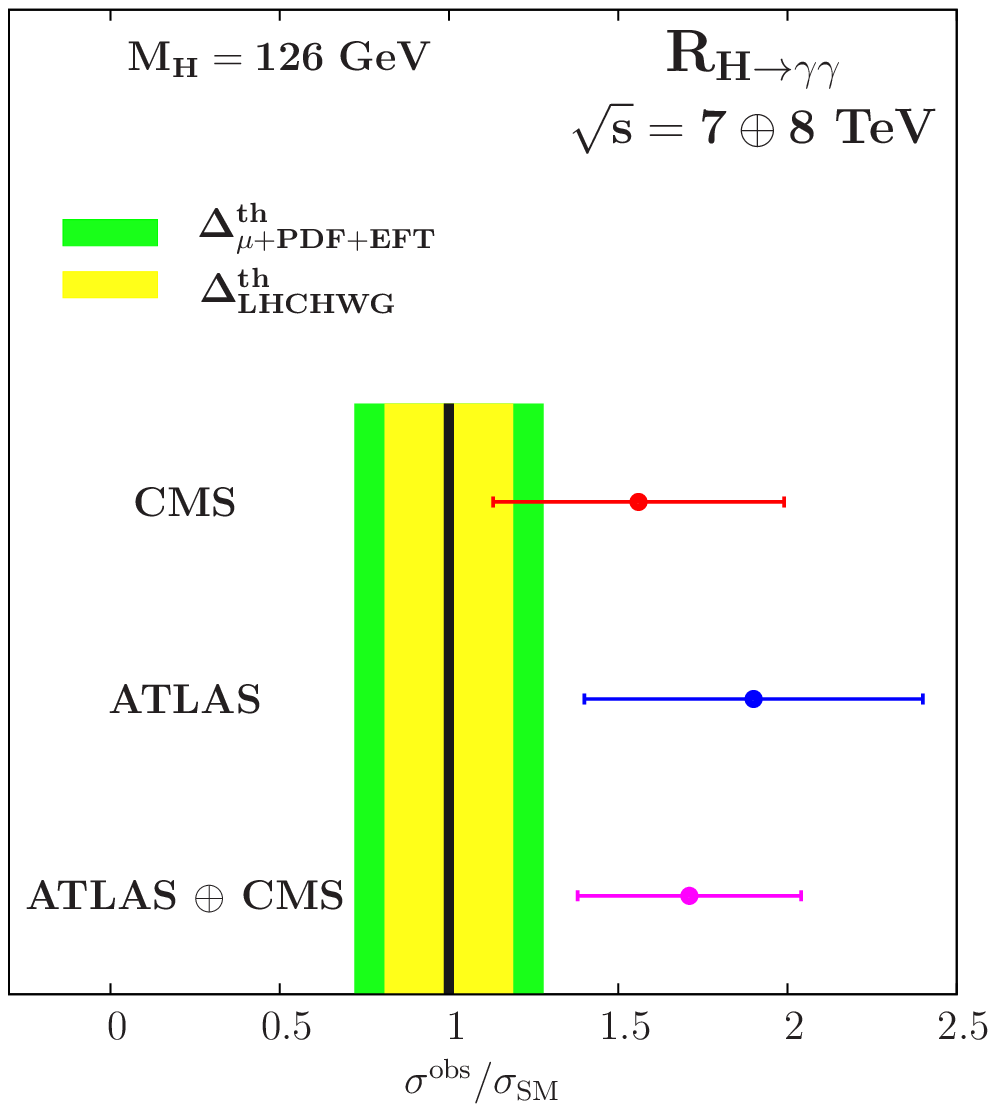,scale=0.73}
\hspace*{5mm}
\epsfig{file=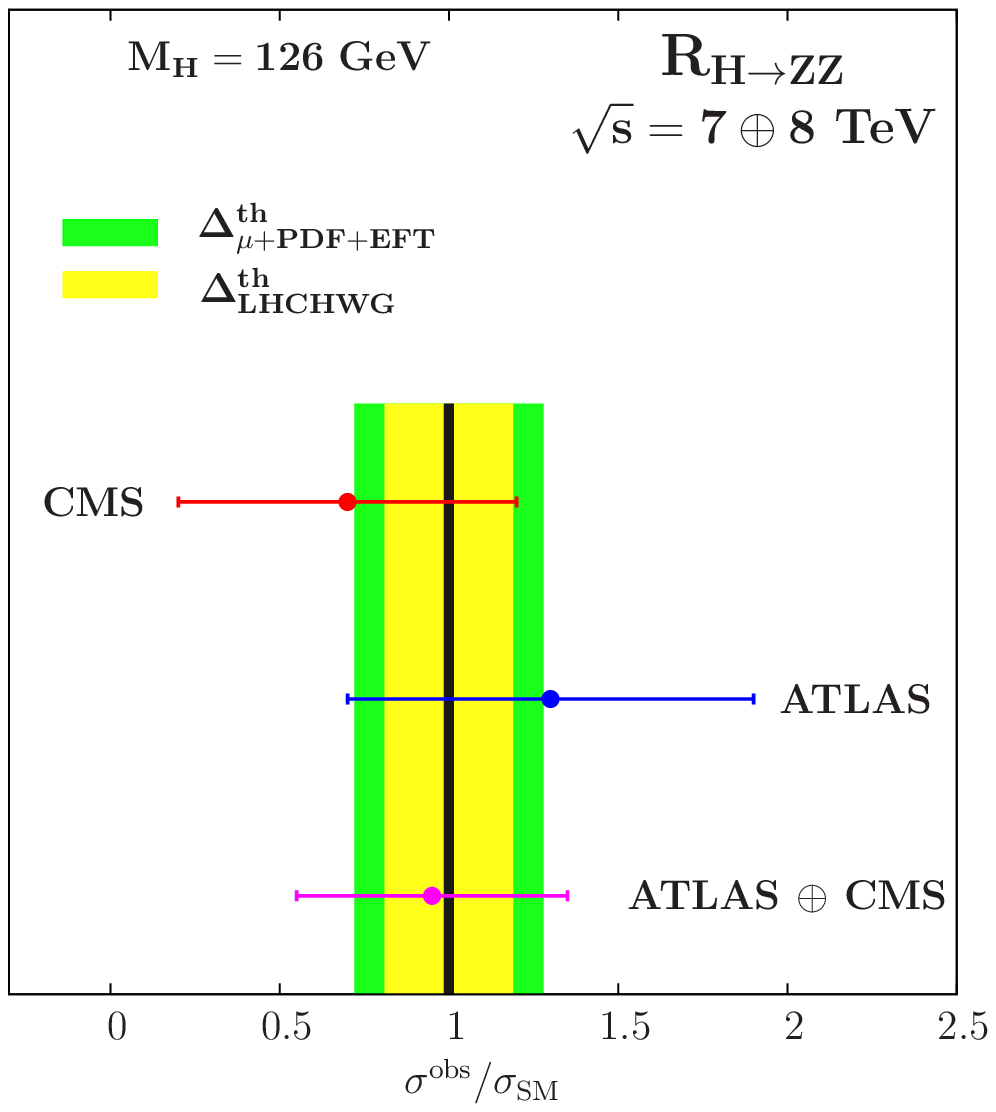,scale=0.73}
\end{center}
\vspace*{-5mm}
\caption[]{\small The value of $R_{XX}$ for the $H\to \gamma \gamma$
  and  $ZZ$ final states given by the ATLAS and CMS Collaborations, as
  well as their combination, compared to the theoretical uncertainty
  bands.}
\label{muhat}
\end{figure}

This is what is done in Fig.~\ref{muhat} for a Higgs mass of 126
GeV. First we combine the theory predictions for the $gg\!\to\! H$
cross section and their uncertainties at $\sqrt s\!=\!7$ and 8 TeV
and, because the integrated luminosity in both the 2011 and 2012 data
samples is approximately the same, we simply perform an average. We
then add all the theoretical uncertainties in two possible scenarios:
$i)$ only the scale and PDF uncertainties as advocated by the LHCHWG
and which leads to $\Delta^{\rm th}_{\rm LHCHWG} \sigma \! \approx \!
\pm 19\%$, $ii)$ add linearly to the previous result the EFT
uncertainty leading to a total of $\Delta^{\rm th}_{\rm \mu + PDF +
  EFT} \sigma\! \approx \! 28\%$.

In Fig.~\ref{muhat}, the green and yellow bands represent these two
possibilities for the total uncertainty. These bands are compared with
the experimental measurements (which again we identify as a first
approximation with the optimal value of the strength modifier $\hat
\mu$) for the normalised production rates in the two
channels\footnote{We do not include the $gg\to H\to WW$ channel as
  first, it has not been yet analysed fully by the ATLAS Collaboration
  and second, the cross section in this case is broken into 0, 1 and 2
  jet bins and this introduces an additional uncertainty due to the
  jet veto which can be significant \cite{scale-Hj}.},
$R_{\gamma\gamma}$ and $R_{ZZ}$ of the ATLAS and CMS Collaborations,
as well as the ATLAS and CMS combination, given in
eq.~(\ref{masterequation}).

It is clear that including the theoretical uncertainty helps to reduce
the discrepancy between the experimental and theoretical values in the
$H\to \gamma \gamma$ channel, while keeping the accord between the
data and the SM prediction in the $H\to ZZ$ channel. In the approach
where the scale, PDF and EFT uncertainty are added linearly, one would
obtain in the $H\to \gamma \gamma$ channel deviations with a
significance of $0.7\sigma, 1.24\sigma$ and $1.3\sigma$ for the CMS,
ATLAS and ATLAS$\oplus$CMS results, respectively.

One should finally comment on the optimal value of the strength
modifier when all channels are combined, $\hat \mu_{\rm tot}$, given
by the ATLAS and CMS Collaborations when the 2011 and 2012 data are
added: $\hat \mu_{\rm tot}=1.2 \pm 0.3$ and $\hat \mu_{\rm tot}=0.8
\pm 0.22$, respectively\footnote{As can be seen, there is an apparent
  deficit in the CMS Higgs cross sections despite the apparent excess
  in $H \to \gamma \gamma$ rate. If the latter should turn out to be
  due to an upward statistical fluctuation and the global deficit
  should remain, the situation can be very easily accomodated by the
  20--30\% theoretical uncertainty in the normalisation of
  $\sigma(gg\to H)$, that we have been discussing here.}. If this
parameter is to be viewed as a cross section measurement, it would
mean that the  $gg\to H$ is already  ``measured" to better than
$\approx 25\%$ (since all channels analysed by the two collaborations
are initiated by $gg \to H$, except for the $H\to b\bar b$ channel for
which the sensitivity is still rather low). This total error, which
should be in principle largely due to the presently limited
statistics, is of the same order of the theory uncertainty in the best
case. We believe that this ``paradox" will be resolved if the approach
that we advocate, that is comparing the data for the cross sections
including only the experimental uncertainties to the theoretical
prediction with the uncertainty bands.\smallskip

In conclusion, we have first recalled that there are substantial
theoretical uncertainties in the cross section for the dominant Higgs
production channel at the LHC, gluon--gluon fusion, stemming from the
scale dependence, the parton distribution functions and the use of an
effective field theory approach to evaluate some higher order
corrections. They are about 10\% each and if they are combined
according to the LHCHWG, they reach the level of $30\%$ when the EFT
uncertainty is also included. However, in the experimental analyses,
these theoretical uncertainties in $\sigma(gg\to H)$ are treated as
nuisance parameters rather than a bias. As they are still individually
smaller than the experimental (statistical) errors, the net result is
as if they had not been included in the total errors given by the
ATLAS and CMS Collaborations. If the experimental results for the
production cross sections times decay  branching ratios in the various
analysed channels are confronted with the theoretical prediction,
including the theoretical uncertainty band, added linearly on top of
the experimental error the discrepancy between the measurements and
the prediction becomes smaller. This is particularly the case for
$\sigma(gg\to H)\times {\rm BR}(H \to \gamma \gamma)$, where the
$\approx 2\sigma$ discrepancy with the SM prediction reduces to the
level of $\approx 1 \sigma$ if the 30\% theory uncertainty is properly
considered.\bigskip

\noindent {\bf Acknowledgements}: A.D. and R.M.G. thank the CERN
theory division for its hospitality during which this project was
completed. R.M.G. acknowledges the project SR/S2/JCB64 DST (India) and
J.B. acknowledges the support from the Deutsche Forschungsgemeinschaft
via the Sonder-forschungsbereich/Transregio SFB/TR-9 Computational
Particle Physics. We thank Marco Battaglia for a discussion on the
data.

\end{document}